\documentclass[a4paper]{article}

\usepackage{is22}
\usepackage{url}
\usepackage{cite}
\usepackage{hyperref}
\usepackage{multirow}
\usepackage{color,soul}

\DeclareMathOperator{\snri}{SNRi}

\newcommand{\ReduceSpaceUnderFigure}{\vspace{-8pt}}
\newcommand{\ReduceSpaceUnderTable}{\vspace{-8pt}}

\title{SNRi Target Training for Joint Speech Enhancement and Recognition}
\name{Yuma Koizumi,
Shigeki Karita,
Arun Narayanan,
Sankaran Panchapagesan,
Michiel Bacchiani}
\address{Google Research}
\email{\{koizumiyuma,karita,arunnt,panchi,michiel\}@google.com}

\begin{document}

\maketitle

\begin{abstract}
Speech enhancement (SE) is used as a frontend in speech applications including automatic speech recognition (ASR) and telecommunication. A difficulty in using the SE frontend is that the appropriate noise reduction level differs depending on applications and/or noise characteristics. In this study, we propose ``{\it signal-to-noise ratio improvement (SNRi) target training}''; the SE frontend is trained to output a signal whose SNRi is controlled by an auxiliary scalar input. In joint training with a backend, the target SNRi value is estimated by an auxiliary network. By training all networks to minimize the backend task loss, we can estimate the appropriate noise reduction level for each noisy input in a data-driven scheme. Our experiments showed that the SNRi target training enables control of the output SNRi. In addition, the proposed joint training relatively reduces word error rate by 4.0\% and 5.7\% compared to a Conformer-based standard ASR model and conventional SE-ASR joint training model, respectively. Furthermore, by analyzing the predicted target SNRi, we observed the jointly trained network automatically controls the target SNRi according to noise characteristics. Audio demos are available in our demo page\footnote{\url{google.github.io/df-conformer/snri_target/}}.
\end{abstract}

\section{Introduction}
\label{sec:intro}

Speech enhancement (SE) is the task of recovering target speech from a noisy signal~\cite{dlwang_2018}. 
Single-channel SE is an indispensable frontend in most speech tasks; for example, improving speech intelligibility for telecommunication~\cite{Moore2003,Wang2008,DnsIcassp2021}, reducing noise for automatic speech recognition (ASR)~\cite{fujimoto_2004,Weninger_2015,hakan_2015,kinoshita_2020,espnetse_2021} systems, and estimating spatial covariance matrix for multi-channel SE~\cite{Erdogan+2016,Higuchi2018}.

A difficulty in using SE frontend is that the appropriate noise reduction level is different depending on the applications and/or noise characteristics. For example, maximizing signal-to-noise ratio (SNR) improvement does not necessarily lead to better ASR performance~\cite{espnetse_2021} and perceptual speech quality~\cite{Koizumi2018}. This is likely due to the distortions introduced by non-linear processing in single-channel SE such as time-frequency (TF) masking. While typical single-channel SE frontends aim to perfectly remove noise, in practice they cause artifacts in the resulting denoised speech.

One strategy to solve this problem is restricting SE performance~\cite{Koizumi2018,Cohen_2002,Vincent_2010,Arun_icassp_2014,Q_Wang_Eusipco_2021,Menne_icassp_2019,Pandey_2021,Sato2022,Iwamoto2022}. Earlier SE studies limit noise suppression in non-speech TF bins by flooring~\cite{Koizumi2018,Cohen_2002}, smoothing~\cite{Koizumi2018,Vincent_2010}, and scaling~\cite{Arun_icassp_2014,Q_Wang_Eusipco_2021} the estimated TF mask in the short-time Fourier transform (STFT) domain. Since recent time-domain SE~\cite{convtasnet,Kavalerov2019,Pariente2020,Scott2020,Koizumi_waspaa_2021} cannot control noise reduction level by manipulating estimated masks, the observed and/or estimated noise signals are added to the estimated speech signal~\cite{Sato2022,Iwamoto2022}. This strategy is certainly effective, and therefore, it would be worthwhile to extend this approach to an more interpretable form, optimized in data-driven manner.

\begin{figure}[t]
  \centering
\includegraphics[width=\linewidth,clip]{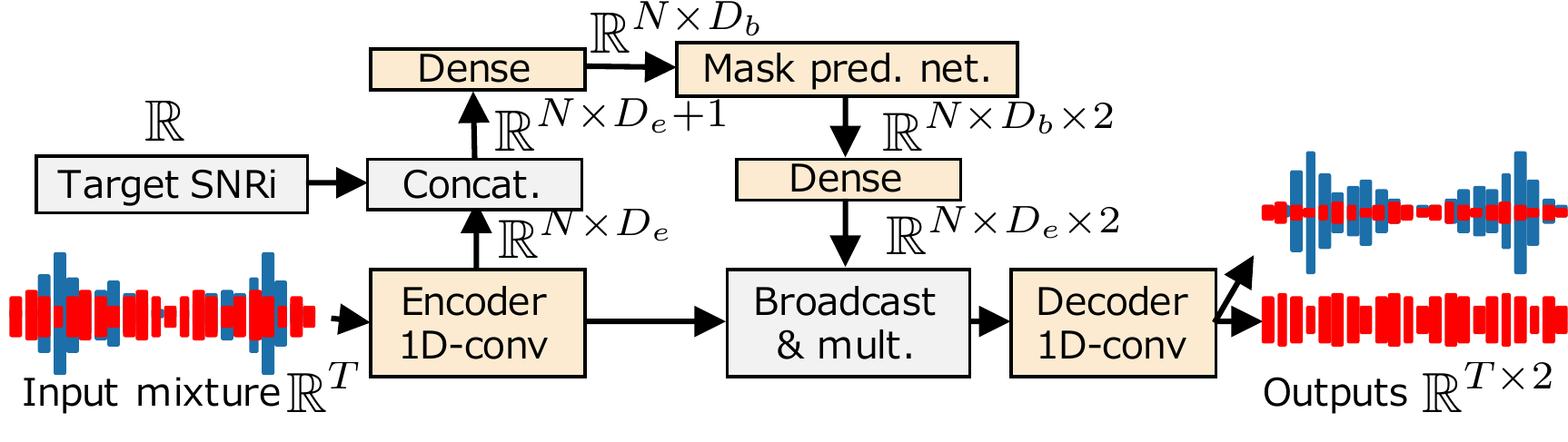} 
  \vspace{-20pt}
  \caption{Network architecture of SNRi-Net trained by SNRi target training. $N$, $D_e$, and $D_b$ mean number of time frames, encoder/decoder basis, and bottleneck feature, respectively.}
  \label{fig:snri_target_training}
  \ReduceSpaceUnderFigure
\end{figure}

\begin{figure}[t]
  \centering
\includegraphics[width=\linewidth,clip]{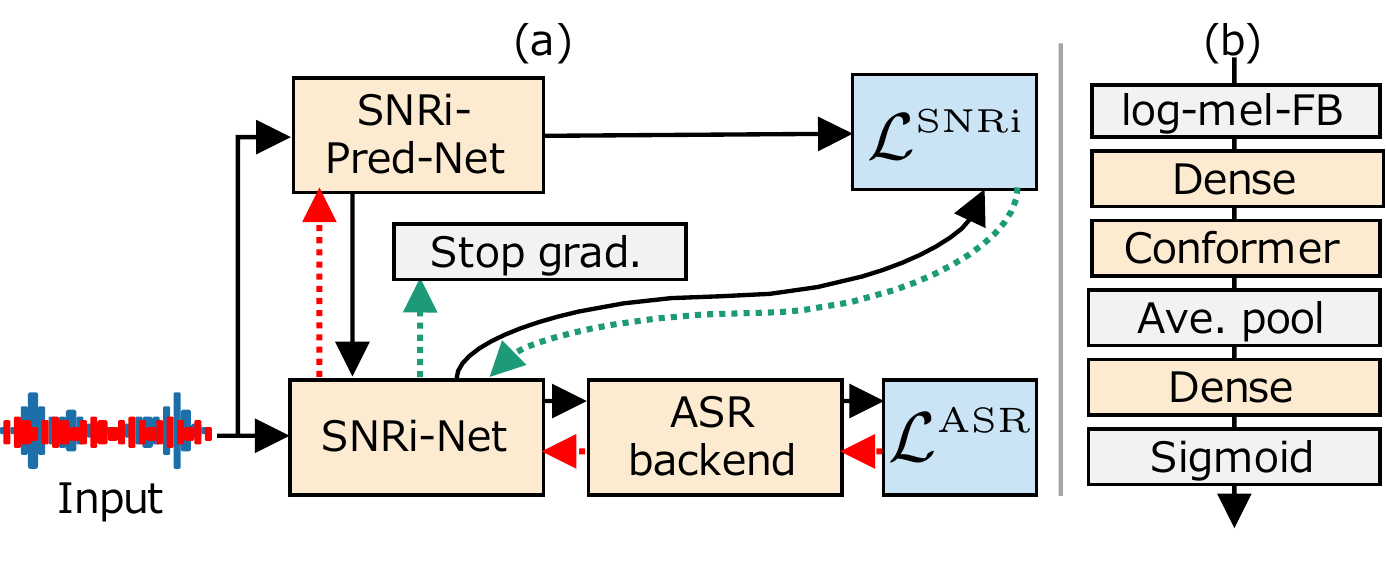} 
  \vspace{-20pt}
  \caption{(a) Overview of proposed joint training of SNRi-Net and ASR backend. (b) Network architecture of SNRi-Pred-Net. Black solid lines show variable flow in forward-propagation, and red/green dotted lines show gradient flow in back-propagation.}
  \label{fig:asr_joint_training}
  \ReduceSpaceUnderFigure
\end{figure}

In this paper, we address the following question: \textit{How much signal-to-noise ratio improvement (SNRi) is required in each task for a given noisy input?}
To provide interpretability to the SE frontend, we propose a new framework named as \textit{SNRi target training}. The SE frontend uses an auxiliary scalar input, which represents the target SNRi of the output signal. Instead of optimizing the enhancement to maximize the SNR, its goal is altered to produce an output signal with the specified target SNRi.
We call this SE model, ``\textit{SNRi-Net}''. A block diagram of SNRi-Net is shown in Fig.~\ref{fig:snri_target_training}.
As a specific use case, we adopt ASR as the backend task to evaluate the merit of SNRi-Net. 
In joint training, the target SNRi value is estimated by an auxiliary network, which we call, \textit{SNRi-Pred-Net}. All networks are trained to minimize ASR loss in an end-to-end manner as shown in Fig.~\ref{fig:asr_joint_training}. This way, we can estimate the appropriate noise reduction level for each noisy input in a data-driven manner.

Experiments show that our SNRi target training enables control of the SNRi more accurately than post-mixing of separate signals. For evaluating the proposed method as the SE frontend, our ASR system was compared with a Conformer-based standard ASR model and conventional SE-ASR joint training model.
We used noisy datasets with and without reverberation, and the proposed method achieved the WER reduction by 12.5\% (dry) and 4.0\% (reverberant) from the standard ASR model, and 1.5\%  (dry) and 5.7\% (reverberant) from the SE-ASR joint model, respectively.
Furthermore, we observed that the predicted target SNRi was controlled dynamically according to noise characteristics.
Audio demos are available in our demo page${}^{1}$.

\section{Conventional Method}
\label{sec:conventional_se}

Let the $T$-sample time-domain observation $\bm{x} \in \mathbb{R}^T$ be a mixture of a target speech $\bm{s}$ and noise $\bm{n}$, such that, $\bm{x} = \bm{s} + \bm{n}$. The goal of standard SE is to recover $\bm{s}$ from $\bm{x}$. A popular strategy in supervised SE is the time-domain mask-based method~\cite{convtasnet,Kavalerov2019,Pariente2020,Scott2020,Koizumi_waspaa_2021}. As an implementation example, \cite{Koizumi_waspaa_2021} estimates masks for separating speech and noise by a mask prediction network and applies it to the representation of $\bm{x}$ encoded by an encoder. The estimated signals $\bm{y} \in \mathbb{R}^{T \times 2}$ are then re-synthesized using a decoder. Here, $\bm{y}_{:, 1}$ and $\bm{y}_{:, 2}$ are the estimates of speech and noise, respectively. Then, a mixture consistency projection layer~\cite{ScottIcassp2020} is applied to ensure the sum of $\bm{y}_{:, 1}$ and $\bm{y}_{:, 2}$ equals $\bm{x}$:
\begin{align}
\bm{y}_{:, 1} \gets \bm{y}_{:, 1} + \zeta \bm{e}, \quad
\bm{y}_{:, 2} \gets \bm{y}_{:, 2} + (1 - \zeta) \bm{e},
\end{align}
where $\bm{e} = \bm{x} - (\bm{y}_{:, 1} + \bm{y}_{:, 2})$, and $\zeta \in [0, 1]$ is a tunable hyperparameter. The negative thresholded SNR~\cite{Scott2020} is used as the loss:
\begin{align}
\mathcal{L}^{\mbox{\tiny{SE}}} &= \alpha \mathcal{L}^{\mbox{\tiny{SNR}}}_{s, y_{:,1}} + (1-\alpha) \mathcal{L}^{\mbox{\tiny{SNR}}}_{n, y_{:,2}},\\
\mathcal{L}^{\mbox{\tiny{SNR}}}_{a,b} &= -10 \log_{10}(\lVert \bm{a}\rVert^2 / (\lVert \bm{a} -  \bm{b}\rVert^2 + \tau\lVert \bm{a}\rVert^2)),
\end{align}
where $\lVert \cdot \rVert$ is $\ell_2$ norm and $\tau = 10^{-3}$ is a soft threshold that clamps the loss at 30 dB~\cite{Scott2020} and $\alpha \in [0, 1]$ is a tunable hyperparameter~\cite{Scott2020}.

The SE fronend can be joined with a backend such as ASR~\cite{kinoshita_2020,espnetse_2021}.
In a typical framework, the estimated speech $\bm{y}_{:, 1}$ is fed to the backend, then both frontend and backed are jointly fine-tuned by minimizing a loss function.
In noise robust ASR tasks, the loss function $\mathcal{L}$ can be a weighted sum of the ASR loss $\mathcal{L}^{\mbox{\tiny{ASR}}}$ and $\mathcal{L}^{\mbox{\tiny{SE}}}$ as
\begin{align}
    \mathcal{L} = \mathcal{L}^{\mbox{\tiny{ASR}}} + \gamma \mathcal{L}^{\mbox{\tiny{SE}}},
\end{align}
where $\gamma \ge 0$ is a tunable hyperparameter.

In practice, $\bm{y}_{:, 1}$ includes artifacts in speech. Such distortions cause degradation of the backend tasks such as ASR. To reduce distortion, several post-processing methods have been proposed~\cite{Arun_icassp_2014,Q_Wang_Eusipco_2021,Menne_icassp_2019,Pandey_2021,Sato2022,Iwamoto2022}. For time-domain speech separation~\cite{Sato2022} and SE~\cite{Iwamoto2022}, a possible strategy is to add $\bm{y}_{:, 2}$\footnote{In conventional studies~\cite{Sato2022,Iwamoto2022}, $\bm{x}$ is added instead of $\bm{y}_{:, 2}$, which is the same as in Eq.~(\ref{eq:post_mixing}) except for the constant multiplication. To clarify the relationship with SNRi, we formulate the post-mixing using $\bm{y}_{:, 2}$.}. Since SE estimates of clean speech $\bm{y}_{:,1}$ and noise $\bm{y}_{:,2}$, we can control noise reduction level as
\begin{align}
    \bm{y} = \bm{y}_{:, 1} + w \bm{y}_{:, 2},
    \label{eq:post_mixing}
\end{align}
where $w \in [0, 1]$. By assuming that the SE module perfectly separate speech and noise, SNRi of $\bm{y}$ can be controlled as $w = 10^{- \lambda / 20}$, where $\lambda$ is a target SNRi scalar.

\section{Proposed Method}
\subsection{SNRi target training}
\label{sec:propose_se}

In contrast the conventional SE, the goal of SNRi target training is to control SNRi of the SE output $\bm{y}_{:, 1}$ according to $\lambda$. To achieve this, $\lambda$ is also input to the mask prediction network as an auxiliary variable. Specifically, we concatenate $\lambda$ to the encoder output in the feature dimension as shown in Fig.\,\ref{fig:snri_target_training}. We call this network as \textit{SNRi-Net}. After applying a mixture consistency projection layer~\cite{ScottIcassp2020}, the loss value is calculated as the squared-error between the target SNRi, $\lambda$, and SNRi of the output signal as:
\begin{align}
    \mathcal{L}^{\mbox{\tiny{SNRi}}} &= \lvert \lambda - \snri \rvert^2 + \beta \mathcal{L}^{\mbox{\tiny{SAR}}},\\
    \snri &= 10 \log_{10} \left( \frac{\lVert \bm{s} \rVert ^2}{\lVert \bm{y}_{:, 1} \!-\! \bm{s} \rVert ^2} \right) - 10 \log_{10} \left( \frac{\lVert \bm{s} \rVert ^2}{\lVert \bm{n} \rVert ^2}  \right),\\
    \mathcal{L}^{\mbox{\tiny{SAR}}} &= -10 \log_{10}(\lVert \bm{s}\rVert^2 / (\lVert \bm{e}_{\mbox{\tiny{artif}}}\rVert^2 + \tau\lVert \bm{s}\rVert^2)),
\end{align}
where $\beta \in [0, 1]$ is a weight parameter and $\mathcal{L}^{\mbox{\tiny{SAR}}}$ is the negative thresholded source-to-artifact ratio (SAR) which reduces artifacts in the output signal based on the SAR~\cite{sisnr}. $\bm{e}_{\mbox{\tiny{artif}}}$ are the artifacts in the output signal which can be obtained by decomposing the residual noise $\bm{y}_{:, 1} - \bm{s} = \bm{e}_{\mbox{\tiny{interf}}} + \bm{e}_{\mbox{\tiny{artif}}}$ where $\bm{e}_{\mbox{\tiny{interf}}}$ is the orthogonal projection of the residual noise onto the subspace spanned by both $\bm{s}$ and $\bm{n}$~\cite{sisnr}.

\subsection{Joint training of SNRi-Net and ASR backend}
\label{sec:propose_asr}

SNRi-Net can also be jointly trained with a backend task, like ASR. The loss function for joint-training then becomes:
\begin{align}
    \mathcal{L} = \mathcal{L}^{\mbox{\tiny{ASR}}} + \eta \mathcal{L}^{\mbox{\tiny{SNRi}}},
    \label{eq:joint_proposed}
\end{align}
where, $\eta \ge 0$ is a tunable hyperparameter.
The problem of the joint network is that the appropriate target SNRi $\lambda$ is unknown for each task and/or input. To address this, we propose using an auxiliary network called as \textit{SNRi-Pred-Net}, which predicts a $\lambda$ as $\hat{\lambda}$ that minimizes $\mathcal{L}^{\mbox{\tiny{ASR}}}$.
Fig.\,\ref{fig:asr_joint_training}\,(a) shows the overview of the proposed joint training framework. The architecture of SNRi-Pred-Net used in this study is shown in Fig.\,\ref{fig:asr_joint_training}\,(b).

Log-mel filterbank is used as inputs by SNRi-Pred-Net. The input features are first passed to a dense layer for computing bottleneck features, which are then processed by Conformer blocks ~\cite{conformer}. 
We apply average-pooling across time to the the output of the Conformer to obtain a single, vector representation for the inputs. A second dense layer followed by a sigmoid function coverts the vector to a scalar, which is the predicted target SNRi. A scaling operation is also used to restrict the predicted target SNRi to the range $[\lambda_{\min},\lambda_{\max}]$. Here, $\lambda_{\min}$ and $\lambda_{\max}$ are hyperparameters that represents the minimum and maximum values of target SNRi $\hat{\lambda}$.

To enhance the noisy signal, $\bm{x}$ and $\hat{\lambda}$ are passed to SNRi-Net, and its speech output $\bm{y}_{:,1}$ is passed to the ASR backend. 
We train the joint network by minimizing Eq. (\ref{eq:joint_proposed}). That is, the ASR model, SNRi-Net and SNRi-Pred-Net are all trained to minimize ASR loss, $\mathcal{L}^{\mbox{\tiny ASR}}$. Importantly, only SNRi-Net uses the SE loss, $\mathcal{L}^{\mbox{\tiny{SNRi}}}$;
we stop the gradient of $\mathcal{L}^{\mbox{\tiny{SNRi}}}$ from optimizing SNRi-Pred-Net. Since the predicted $\lambda$ is the prediction target for SNRi-Net, if the SE loss is back-propagated to SNRi-Pred-Net, this network will be optimized to output $\lambda$ that makes the task of SNRi-Net easy, but sub-optimal for ASR.

\section{Experiment}
\label{sec:experiment}

\subsection{Experimental settings}
\label{sec:experiment_setup}

\noindent
{\bf Dataset:}
We used the same training dataset as \cite{Arun_ASRU_2021} with and without reverberation.
The dataset without reverberation is called ``\texttt{Dry}'' and with reverberation as ``\texttt{Reverb}''.
This dataset includes 4,249 hours of clean speech which consists of 281k utterances from the LibriSpeech~\cite{librispeech} training set and 1,916k utterances from an internal dataset. The noisy utterances were generated using a room simulator~\cite{Kim_RIRsim_2017}, with SNR from -10 dB to 30 dB. Noise is sampled from internally collected noise snippets that simulate conditions like cafe, kitchen, and cars, and freely available noise sources from \href{https://www.gettyimages.com/about-music}{\texttt{Getty}}~\cite{Getty} and \href{https://youtube.com/audiolibrary}{\texttt{YouTube Audio Library}}~\cite{YtAudioLib}. For \texttt{Reverb}, the room configurations have reverberation times (RT60) ranging from 0 ms to 900 ms. For \texttt{Dry}, RT60 is always 0 ms. We generated multiple copies of the data under different mixing conditions in order to model enough combinations of clean speech, background noise, and room-configuration. The training dataset includes 39.6M noisy samples (55,027 hours) in total.

For the validation and test datasets, we used utterances from the LibriSpeech~\cite{librispeech} dev and test sets. The noisy utterances were generated using the same manner as the training dataset with SNR from -5 dB to 20 dB, but with noise sources that are held out from the training dataset. We generated ``\texttt{Dry}'' and ``\texttt{Reverb}'' dev and test sets using the same T60 condition as the training set. The test dataset consists of 2,620 utterances and the average SNR of the input signals was 7.5 dB.

\vspace{2pt}
\noindent
{\bf Models and hyper-parameters:}
As SNRi-Net, we used the same architecture of DF-Confromer-8~\cite{Koizumi_waspaa_2021} except for concatenating the target SNRi. All hyper-parameters were the same as that used in \cite{Koizumi_waspaa_2021}; the number of Conformer layers was 8, $D_e = 256$, $D_b = 216$, and the window and hop sizes of filterbanks were 2.5 ms and 1.25 m. Other hyper-parameters were decided based on the WER on the validation dataset as $\beta = 0.01$, $\lambda_{\min} = 0.0$ and $\lambda_{\max} = 20.0$.

We use the large size Conformer Transducer ASR backend referred as Conformer(L)~\cite{conformer} that has 17 Conformer blocks of 512-dim 8-head dot-product attention, 512-dim 32-frame kernel 1d convolution, 2048-dim feedforward module, and a 640-dim long short-term memory (LSTM) decoder. The input feature was 80-dim filterbank features computed from a 25 ms window with a stride of 10 ms. We use SpecAugment~\cite{SpecAugment} with mask parameter ($F = 27$), and 10 time-masks with maximum time-mask ratio ($p_S = 0.05$). All transcriptions were tokenized with a word piece model with a 1,024 vocabulary built from LibriSpeech 960h. The ASR loss $\mathcal{L}^{\mbox{\tiny{ASR}}}$ was the recurrent neural network transducer (RNN-T) loss~\cite{rnnt,Karita_is_2021}.

SNRi-Pred-Net had 2-Conformer-blocks where each block has 128-dim 6-head dot-product attention, 128-dim 5-frame kernel 1d convolution, and 512-dim feedforward module. The parameter of mel-filterbank was the same as the ASR backend.

SNRi-Net and ASR backend were pre-trained for 200k steps individually; SNRi-Net was pre-trained to minimize $\mathcal{L}^{\mbox{\tiny{SNRi}}}$ with the training dataset, and the ASR backend was pre-trained to minimize $\mathcal{L}^{\mbox{\tiny{ASR}}}$ using both clean and noisy speech in the training dataset. While pre-training of SNRi-Net, $\hat{\lambda}$ was randomly drawn from the uniform distribution $\mathcal{U}(\lambda_{\min},\lambda_{\max})$.

The joint network was fine-tuned for an additional 100k steps to minimize Eq.\,(\ref{eq:joint_proposed}) where $\eta = 0.01$ was determined based on WER on the validation dataset. In the fine-tune stage, we skipped SNRi-Net and SNRi-Pred-Net with a probability of 5\% as a multi-condition learning strategy~\cite{kinoshita_2020}. In addition, we used a random $\hat{\lambda}$ drawn from $\mathcal{U}(\lambda_{\min},\lambda_{\max})$ instead of the predicted one with 25\% probability. All training used the Adam optimizer~\cite{kingma2014adam} with the same setting as~\cite{Koizumi_waspaa_2021} except for using 1/10 learning rate in the fine-tuning stage, and 128 Google TPUv3 cores with a global batch size of 512.

\subsection{Evaluation for control accuracy of output SNRi}
\label{sec:experiment_snri}

We compared SNRi target training with post-mixing based SNRi control in Eq.\,(\ref{eq:post_mixing}).
We used DF-Conformer-8~\cite{Koizumi_waspaa_2021} with hyper-parameters set to the same values as~\cite{Koizumi_waspaa_2021}, i.e. $\alpha = 0.8$ and $\zeta = 0.5$. We tested both methods with two input SNR conditions; SNR of all test samples were adjusted to -5 dB and 5 dB. The target SNRi were 3 dB, 6 dB, 9 dB, and 12 dB. 

\begin{figure}[t]
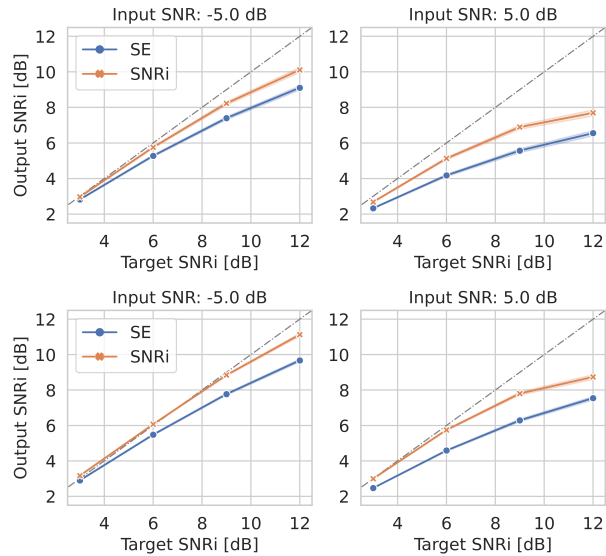

  \centering
  \includegraphics[width=\linewidth,clip]{./snri_eval_result_dry.pdf} 
  \includegraphics[width=\linewidth,clip]{./snri_eval_result.pdf} 
  \vspace{-16pt}
  \caption{Results of experiment on controlling SNRi of output signal. Top and bottom figures are results on  \texttt{Dry} and \texttt{Reverb} test sets, respectively. Solid lines and colored are show mean and 99 \% confidence intervals, respectively. Legends SE and SNRi mean SE+post-mixing and SNRi-Net, respectively.}
  \label{fig:snri_eval_result}
  \ReduceSpaceUnderFigure
\end{figure}

Figure\,\ref{fig:snri_eval_result} shows the experimental results and audio examples of SNRi-Net are available in our demo page$^{1}$.
SNRi-Net trained by SNRi target training achieved significantly better control accuracy of output SNRi than supervised SE on both \texttt{Dry} and \texttt{Reverb} test sets. Although the post-mixing-based control assumes that the output signals are perfectly separated, the outputs usually contains separation errors. Such separation errors affects the output SNRi of the post-mixed signal, resulting in the under-separation problem; the output SNRi is always lower than the target SNRi. SNRi target training has succeeded to avoid the under-separation problem by directly controlling output SNRi in 3 dB and 6 dB target SNRi conditions.

Whereas in 9 dB and 12 dB target SNRi conditions, output SNRi of both methods are significantly lower than target SNRi, even though SNRi-Net was still better than the supervised method. It might be due to the performance upper-bound of the base SE network. It is necessary to accurately distinguish speech and noise to output high SNR signals, and the accuracy should correlate on the separation performance of the SE network. Therefore, to achieve higher control performance in high target SNRi conditions, it is necessary to improve the base SE network performance.

\subsection{Evaluation as ASR frontend}
\label{sec:experiment_asr}

\begin{table}[ttt]
\caption{Word error rate on test dataset and number of trainable parameters for each model.}
\label{tab:wer_result}
\centering
\begin{tabular}{ c| c | c c}
\toprule
\multirow{2}{*}{\textbf{Network}} & \multirow{2}{*}{\textbf{\#Params}} 	& \multicolumn{2}{c}{\textbf{WER [\%] ($\downarrow$)}} \\	
& & \texttt{Dry} & \texttt{Reverb} \\
\midrule
Conformer(L) &  118.8M & 10.5 & 14.4 \\	
Conformer(L)+ & 132.3M & 10.7 & 14.9 \\ \midrule
SE+Conformer(L) & 127.4M & 9.3 & 14.7\\ 
SNRi+Conformer(L) &128.3M & {\bf 9.2} & {\bf 13.8} \\	
\bottomrule
\end{tabular}
\ReduceSpaceUnderTable
\end{table}

\begin{figure*}[t]
  \centering
\includegraphics[width=\linewidth,clip]{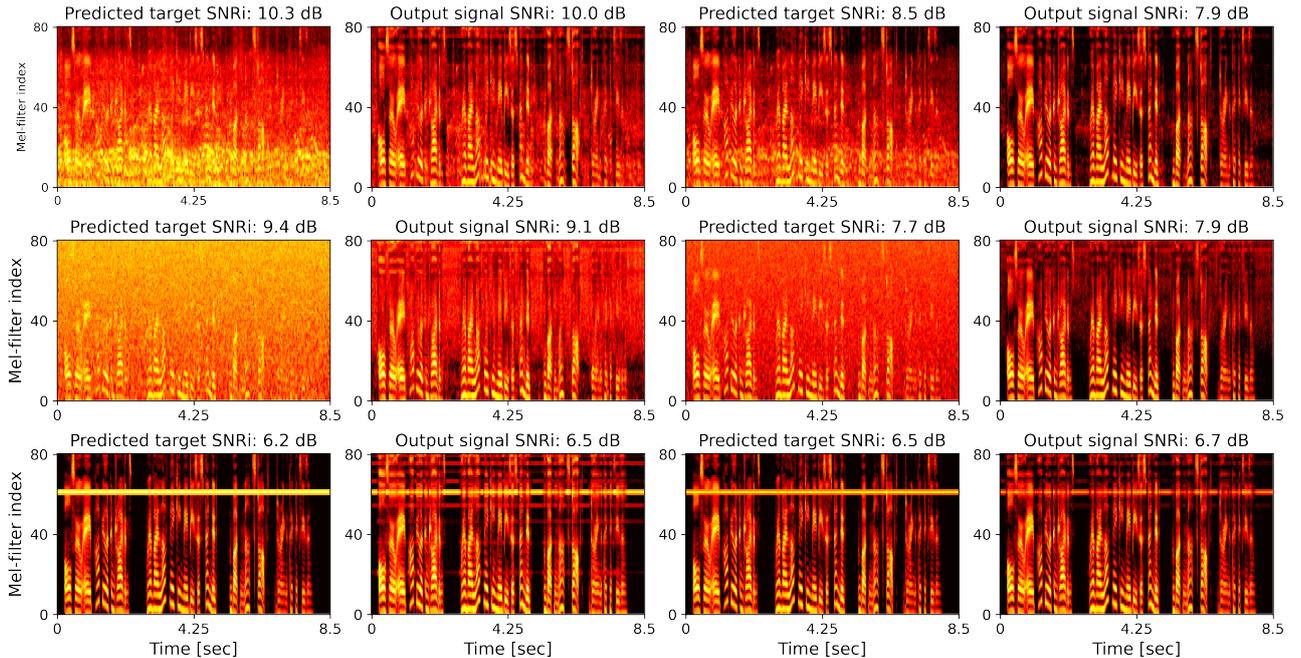} 
  \vspace{-20pt}
  \caption{Log-mel spectrogram examples of input and output of joint trained model. Input SNR of first two column figures are -5.0 dB, and later two figures are 5.0 dB. In the same input SNR figures, left figures are input signal, and right ones are output signal. Noise type of top, middle, and bottom figures are environmental noise, white noise, and 4 kHz sine wave.}
  \label{fig:output_example}
  \ReduceSpaceUnderFigure
\end{figure*}

We compared the ASR performance of the proposed joint-network on WER metrics with two standard ASR models and an SE-ASR joint model.
The first model is Conformer(L)~\cite{conformer}, and the second model is Conformer(L)+ which has 19 Conformer blocks and whose model size is roughly the same as the proposed method\. These models give us the baseline WER of using a large ASR models without SE frontend. These models were trained for 300k steps from scratch.
The third model jointly trains the supervised SE frontend and Conformer(L) ASR model. This model gives us WER of the joint training strategy of an SE frontend and a large ASR backend. The network architecture of the SE frontend was DF-Conformer-8 used in Sec,\,\ref{sec:experiment_snri}. $\gamma = 0.25$ was determined based on WER on the validation dataset. In addition, we skipped the SE frontend with 50\% probability in the fine-tuning stage as a multi-condition learning strategy~\cite{kinoshita_2020}. In this experiment, we excluded Eq.\,(\ref{eq:post_mixing})-based joint models~\cite{Sato2022,Iwamoto2022} for fair comparison because these methods did not fine-tune the joint model.

Table\,\ref{tab:wer_result} shows WER on the test datasets. The proposed method achieved the best WER performance on both datasets; it reduced WER by 12.5\% and 1.5\% compared to Conformer(L) and SE+Conformer(L) for \texttt{Dry}, and 4.0\% and 5.7\% \texttt{Reverb}, respectively. In addition, SE-Conformer(L) increased WER on \texttt{Reverb} compared to Conformer(L), whereas SNRi-Conformer(L) improved WER. From these results, the proposed method consistently improved the ASR accuracy by comparing to merely increasing the ASR backend model size and/or jointly training with a conventional SE frontend.

Although the performance gains in terms of ASR quality from the SE+Conformer(L) method are limited the gains a consistent across conditions. Moreover, the model provides a novel way of controlling the SNRi, thereby providing new insights on how to model SE in a larger system.
Figure\,\ref{fig:output_example} shows log-mel spectrogram examples of the input and output of the proposed joint model. These examples indicate that the predicted SNRi is affected by two factors; input SNR and noise type. By comparing input SNR at -5 dB and 5 dB cases, the predicted SNRi tends to be high when the input SNR is low. This is an intuitive result due to the fact that the larger noise makes ASR more difficult. The other aspect is the noise type; in the case of noise types that only affects a certain frequency, such as a tonal noise, the predicted SNRi tends to be low even if the input SNR is low. As can be seen from the bottom spectrograms, if the majority of the frequency bands are clean, the speech characteristics can be analyzed even if the noise is large. These results show that the proposed method also provides interpretable insights for future development of single-channel SE frontends, e.g. tuning $w$ in Eq.\,(\ref{eq:post_mixing}) using a small scale dataset without joint training~\cite{Iwamoto2022}.

\section{Conclusion}
\label{sec:conclusion}

We proposed ``{\it SNRi target training}''; the SE frontend uses an auxiliary scalar input which represents target SNRi, and enhances the input so that SNRi of the output signal is close to the desired target value.
In joint training with the ASR backend, the target SNRi value was also estimated to minimize the ASR loss.
Experiments showed that SNRi-Net controls the SNRi more accurately than post-mixing of separated signals, and our joint ASR system achieved the best WER on noisy ASR datasets.
Furthermore, by analyzing the output of the joint model, we observed the model automatically controls the target SNRi according to noise characteristics.

The limitation of the proposed method is the requirement of clean speech, in contrast to several conventional joint training strategies that only need noisy speech and its transcription~\cite{kinoshita_2020,Sato2022,eskimez2021human}. Future work will include fine-tuning only SNRi-Pred-Net~\cite{Sato2022} and/or incorporating an unsupervised SE training~\cite{Scott2020}.

\newpage
\bibliographystyle{IEEEtran}
\bibliography{mybib}

\end{document}